\documentclass[10pt, conference, letterpaper]{IEEEtran}
\usepackage{comment}
\usepackage[latin1]{inputenc}
\usepackage[T1]{fontenc}

\def\BibTeX{{\rm B\kern-.05em{\sc i\kern-.025em b}\kern-.08em
    T\kern-.1667em\lower.7ex\hbox{E}\kern-.125emX}}

\usepackage{graphicx}
\usepackage{subfigure}
\usepackage{algorithm}
\usepackage{algorithmicx}
\usepackage{algpseudocode}
\graphicspath{{./figures/}}
\usepackage{amsmath}
\usepackage{url}
\usepackage{array}
\usepackage{multirow}

\begin{document}
\title{PriMe: Per-Flow Network Measurement by Combining SRAM with DRAM}
\author{Zongyi Zhao, Xingang Shi, Xia Yin, Zhiliang Wang\\ zhaozong16@mails.tsinghua.edu.cn shixg@cernet.edu.cn yxia@tsinghua.edu.cn wzl@cernet.edu.cn\\ 
	Tsinghua University}
\maketitle
\begin{abstract}
Network measurement is necessary to obtain an understanding of the network traffic and keep the network healthy. Flow-level measurement is widely used because it provides rich enough information while being resource efficient, in contrast to the traffic-level solutions like SNMP and packet-level solutions like tcpdump. Flow-level measurement requires to maintain a TFR for each active flow, and update the corresponding TFR when a packet arrives. In the scenario of high-speed networks, the processing time for each packet is only several or tens of nanoseconds, and there may be up to millions of concurrent active flows. Two requirements have to be met to be able to do flow-level measurement in this scenario, i.e., the memory size have to be large enough to maintain a TFR for each flow, and the access time of the memory have to be small enough to be able to update the TFR for each packet. Unfortunately, SRAM, which is fast enough, is expensive and there are limited amount available for the measurement tasks, while DRAM, which is large in capacity, is slow and cannot support the per-packet access. In this paper, we proposed a solution, PriMe, by combining SRAM with DRAM, overcoming the resource limitation by exporting TFRs to the DRAM when necessary, while enjoying the fast access of SRAM. We designed a peculiar algorithm to reduce the frequency of exporting TFRs to DRAM, as well as an algorithm to reduce the accesses of DRAM in aggregating the TFRs exported from SRAM. 
\end{abstract}
\section{Overview of PriMe}
PriMe consists of two parts which are in SRAM and DRAM respectively. The new arriving packets are processed by the SRAM part, and the temporary result generated by the SRAM part will be handled by the DRAM part, where traffic report for the network traffic is produced. In the following, we will describe the two parts in detail.

\subsection{SRAM Part}
To guarantee the richness of information, we adopt the IPFIX\cite{trammell_flow_2013} format and focus on the fields of source IP address, destination IP address, source port, destination port, protocol, start timestamp, end timestamp, and packet count. Moreover, source IP address, destination IP address, source port, destination port and protocol are used to identify a flow (they are referred to as flow ID) while start timestamp and end timestamp are used to help the management of on-chip memory, i.e., SRAM.

The SRAM part contains a hash table, $\mathbf{T}$, each bucket of which is to maintain a Temporary Flow Record (TFR). There are $d$ hash functions, $h_1(.), h_2(.), \cdots, h_d(.)$, each of which can hash a flow ID to a bucket of $\mathbf{T}$.

When a packet arrives, the SRAM part will extract the flow ID from the packet, and hash the packet to $d$ different buckets of $\mathbf{T}$. If a TFR with the same flow ID as the packet (i.e., a matching TFR) is found in the $d$ positions, we will update the TFR, i.e., increment the \emph{packet count} field by 1 and write the current system timestamp into the \emph{end timestamp} field. If a matching TFR is not found but an empty bucket is available, we will create a new TFR for this packet and write the current system timestamp into the \emph{end timestamp} field as well. However, if neither a matching TFR nor an empty bucket is found, we will consider evicting an existing TFR from $\mathbf{T}$ to accommodate the new packet. That a TFR has a larger \emph{end timestamp} implies that the flow is active more recently, so we will evict the TFR with the smallest \emph{end timestamp}. Notice that the number of mice flows in a typical network is far greater than the number of elephant flows. In particular, in our CAIDA trace more than half of the flows contains only one packet. So it's very likely that the newly arriving packet corresponds to a mice flow. To encourage that a TFR corresponding to a mice flow is evicted preferentially, we update the \emph{end timestamp} of a newly created TFR as follows:
\begin{eqnarray}
S_1 \gets (S_0 + S)/2
\end{eqnarray}

where $S_1$ is the \emph{end timestamp} of the newly created TFR, $S_0$ is the \emph{end timestamp} of the evicted TFR, and $S$ is the current system timestamp. So even if TFRs in a bucket are evicted constantly, the value of \emph{end timestamp} of the TFRs will increase slowly, and the TFRs in this bucket will obtain higher priority to stay at the hash table. When a TFR with only a packet is evicted, the \emph{end timestamp} will be corrected using the \emph{start timestamp}. The algorithm is shown in Algorithm~\ref{alg: process_packet}.

\begin{algorithm}[ht!]
	\caption{The algorithm of PriMe on the SRAM part}
	\label{alg: process_packet}
	\algrenewcommand\algorithmicwhile{\textbf{when}}
	\begin{algorithmic}[1]
		\State{$p$: the new arriving packet}
		\State{$S$: the current system timestamp}
		\State{$ets$: flow end timestamp}
		\State{$sts$: flow start timestamp}
		\State{$flowID \gets p.flowID, min \gets \infty, pos \gets -1$}
		\For{$i=1$ to $d$}
		\State{$idx\gets h_{i}(flowID)$}
		\If{${\mathbf {T}}[idx].flowID==\text{NULL}$}
		\State{${\mathbf {T}}[idx] \gets (flowID, 1, S)$}
		\State\Return
		\ElsIf {${\mathbf {T}}[idx].flowID == flowID$}
		\State{${\mathbf {T}}[idx].count\gets {\mathbf {T}}[idx].count+1$}
		\State{${\mathbf T}[idx].ets\gets S$}
		\State\Return
		\ElsIf{${\mathbf T}[idx].ets < min$}
		\State{$min \gets {\mathbf T}[idx].ets$}
		\State{$pos \gets idx$}
		\EndIf
		\EndFor
		\State \emph{\# Evict the existing TFR:}
		\If{$\mathbf{T}[pos] == 1$}
		\State{$\mathbf{T}[pos].ets\gets \mathbf{T}[pos].sts$}
		\EndIf
		\State{Export $\mathbf{T}[pos]$ to DRAM}
		\State{$\mathbf{T}[pos].flowID \gets flowID$}
		\State{$\mathbf{T}[pos].count\gets 1$}
		\State{$\mathbf{T}[pos].ets\gets(\mathbf{T}[pos].ets + S$)/2}
	\end{algorithmic}
\end{algorithm}

\begin{figure}
	\centering
	\includegraphics[width=\linewidth]{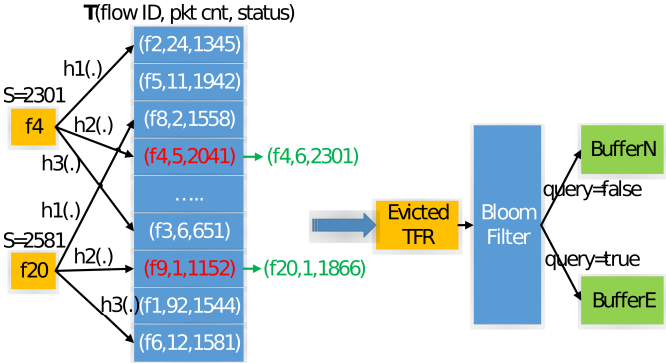}
	\caption{Framework of PriMe's SRAM part}
	\label{fig:framework}
\end{figure}

As shown in Fig.~\ref{fig:framework}, when a packet from flow $f_4$ arrives, we hash it to three different buckets of $\mathbf{T}$. Since a matching TFR exists in $\mathbf{T}$, we increment the packet count of the TFR by 1, and set the \emph{end timestamp} to the current system timestamp, which is denoted by $\mathbf{S}$ in the figure. When a packet from flow $f_{20}$ arrives, since the three buckets corresponding to the flow are not empty, and none of the TFRs are a matching one, we evict the TFR with the smallest \emph{end timestamp}, i.e., the TFR corresponding to $f_9$, and create a TFR for $f_{20}$, set the \emph{end timestamp} to 1866, which is the mean of the \emph{end timestamp} of $f_9$ and the current system timestamp.

Since there are usually multiple TFRs corresponding to an elephant flow, we consider aggregating the TFRs in DRAM. However, as stated above, since many flows have only one packet, thus correspond to only a TFR, it will be a waste of time for DRAM part to aggregate every TFR as aggregating a TFR requires at least a DRAM access. So we consider aggregating the TFRs which are most likely to have other TFRs with the same flow ID. Our idea is that if TFRs with the same flow ID with the current TFR have been seen before, then it is very likely that there will be more TFRs with the same flow ID. So we maintain a bloom filter $\mathbf{B}$, and two buffers, i.e., \emph{BufferE} for existing TFRs and \emph{BufferN} for new TFRs. When a TFR is evicted from $\mathbf{T}$, we query $\mathbf{B}$ using the flow ID. If $\mathbf{B}$ returns a positive response, which means at least one TFR with the same flow ID has been seen, we will put the TFR into \emph{BufferE}. Otherwise, we will put the TFR into \emph{BufferN} and set the corresponding bits in $\mathbf{B}$.

\subsection{DRAM Part}
\begin{figure}[ht!]
	\centering
	\includegraphics[width=\linewidth]{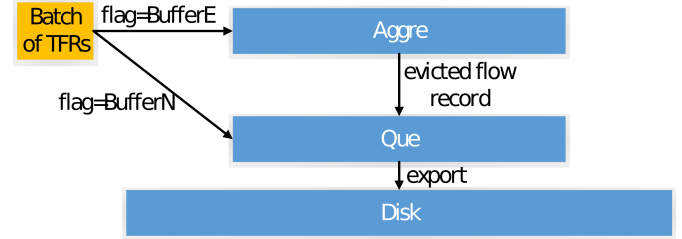}
	\caption{Framework of PriMe's DRAM part}
	\label{fig:drampart}
\end{figure}

As shown in Fig.~\ref{fig:drampart}, the DRAM part of PriMe consists of two components, i.e., an aggregator, \emph{Aggre}, which is a hash table used to aggregate the TFRs exported from the SRAM part, and a queue, \emph{Que}, which is to store the TFRs sequentially.

When either \emph{BufferE} or \emph{BufferN} is full, the buffer will be copied into DRAM through DMA. There is a flag to signal if the batch is from \emph{BufferE} or \emph{BufferN}. If the batch of TFRs is from \emph{BufferN}, which means every TFR of the batch is a new one, we will copy the batch into \emph{Que} directly, so only a read operation and a write operation are needed. Otherwise, if the TFRs are from \emph{BufferE}, which means that at least a TFR with the same flow ID exists for each TFR in the batch, we will hash the TFRs one by one into \emph{Aggre}, trying to aggregate different segments of the same TFR into one. Only when the hash table of \emph{Aggre} is full and collision occurs, the existing TFR is evicted and the new TFR is inserted into \emph{Que}. The evicted TFR will be inserted into \emph{Que}. This process can be described in Algorithm~\ref{alg: DRAM_part_algorithm}.

\begin{algorithm}[ht!]
	\caption{The algorithm of PriMe on the DRAM part}
	\label{alg: DRAM_part_algorithm}
	\algrenewcommand\algorithmicwhile{\textbf{when}}
	\begin{algorithmic}[1]
		\State{ $\Theta$: a batch of TFRs transfered to DRAM}
		\If{$\text{flag}==\text{BufferN}$}
		\State{Copy $\Theta$ into \emph{Que}}
		\State{\Return}
		\EndIf
		\If{$\text{flag}==\text{BufferE}$}
		\For{each $r$ in $\Theta$}
		\State{$r_0=insert\_flow\_record\_into\_aggre(r)$}
		\If{$\text{NULL}\ne r_0$}
		\State{Add $r_0$ into \emph{Que}}
		\EndIf
		\EndFor
		\EndIf
	\end{algorithmic}
\end{algorithm}

In \emph{Aggre}, when a TFR is hashed to a bucket of the table, if a matching TFR is found, the TFR will be merged with it. Otherwise, if there is an existing TFR and the existing one has different flow ID with the current TFR, the existing one will be evicted. This process is described in Algorithm~\ref{alg: insert_flow_record_into_aggre}. Since \emph{Aggre} is in DRAM, we can make it large enough to achieve a small collision rate. So when the size of \emph{BufferE} and \emph{BufferN} are large enough, the time complexity of the DRAM part, including the DRAM accesses and computations, is mainly determined by the number of TFRs inserted into \emph{Aggre}.

\begin{algorithm}[ht!]
	\caption{Insert a TFR into \emph{Aggre}}
	\label{alg: insert_flow_record_into_aggre}
	\algrenewcommand\algorithmicwhile{\textbf{when}}
	\begin{algorithmic}[1]
		\State{$r$: a temporary flow record}
		\State{$ets$: flow end timestamp}
		\Procedure{insert\_flow\_record\_into\_aggre}{$r$}
		\State{$idx\gets hash(r.flowID)$}
		\If{$Aggre[idx].flowID == \text{NULL}$}
			\State{$Aggre[idx]=r$}
		\ElsIf{$Aggre[idx].flowID == r.flowID$}
		\State{$Aggre[idx].count\gets Aggre[idx].count + r.count$}
		\State{$Aggre[idx].ets\gets r.ets$}
		\Else
		\State{$\text{temp}\gets Aggre[idx]$}
		\State{$Aggre[idx]\gets r$}
		\State{\Return temp}
		\EndIf
		\EndProcedure	
	\end{algorithmic}
\end{algorithm}

\section{Evaluation}
In this section, we evaluate the performance of the SRAM part and DRAM part of PriMe separately. For the SRAM part, we take the method used in TurboFlow\cite{sonchack_turboflow:_2018} as a benchmark, i.e., hash a packet to a bucket of the hash table, and evict the existing TFR whenever collision occurs. We implement PriMe and TurboFlow in python to evaluate their performance theoretically. Real network traces from CAIDA and HGC (Hong Kong Global Communication) are used to generate the packets. When calculating the capacity of the hash table, we assume that a single TFR occupies 71 bytes, which is the same as that in NetFlow. The metric used is \emph{eviction rate}, which is defined as $\frac{m}{n}$, where $m$ is the number of TFRs evicted from the hash table in the SRAM part, and $n$ is the total number of packets processed by the switch.

For the DRAM part, our main concern is that there are so many TFRs needing to be processed that the DRAM cannot keep up with the SRAM in the sense of memory access time. Our solution is to process only part of the TRFs exported from the SRAM part while allowing multiple TFRs exist for a single flow. So we evaluate the performance of DRAM part using two metrics: \emph{aggregation rate}, which is defined as $\frac{m_0}{n}$, where $m_0$ is the number of TFRs processed by \emph{Aggre} one by one and $n$ is the total number of packets processed by the switch, and \emph{redundancy}, which is defined as $\frac{k - k_0}{k}$, where $k$ is the number of TFRs finally generated by the DRAM part and $k_0$ is the real number of flows.
 
\subsection{The Performance of SRAM Part}
\begin{figure}[t]
	\centering
	\mbox{
		\subfigure[CAIDA\label{MT-flowrecord}]{\includegraphics[width=0.45\linewidth]{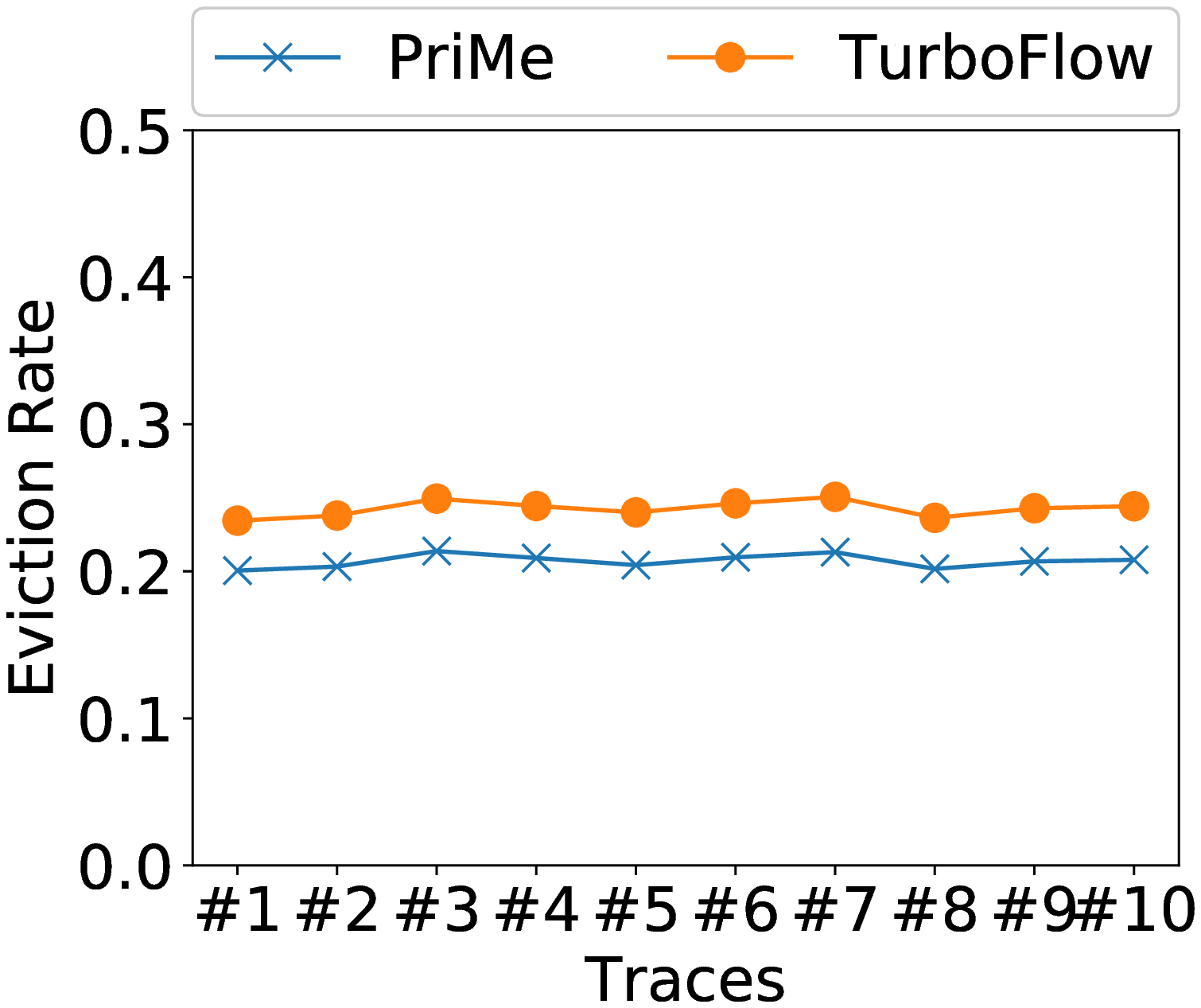}}
		\subfigure[HGC\label{MT-flowsize}]{\includegraphics[width=0.45\linewidth]{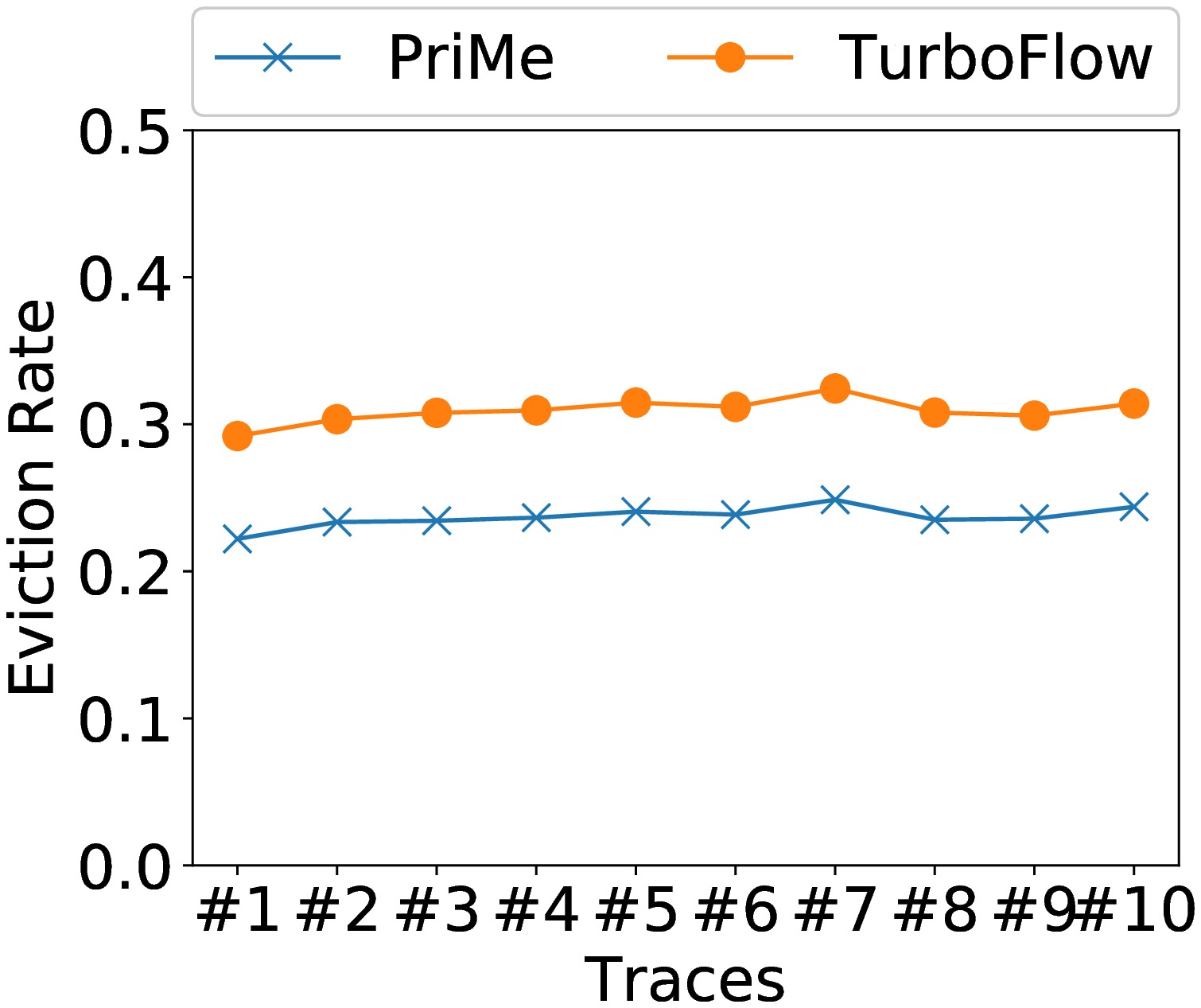}}
	}
	\caption{The \emph{eviction rates} of PriMe and TurboFlow on different trace files from CAIDA and HGC.}
	\label{fig:collision_rates_on_various_trace_files}
\end{figure}

To evaluate the performance of the SRAM part, we set the memory size to 0.5 MB, and run the algorithm using 10 trace files from CAIDA and HGC respectively. From each trace file, we extract and replay 5 million packets. As shown in Fig.~\ref{fig:collision_rates_on_various_trace_files}, PriMe can reduce the \emph{eviction rate} by around 15\% in the CAIDA traces, and around 24\% in the HGC traces. 

\begin{figure}[t]
	\centering
	\mbox{
		\subfigure[CAIDA\label{CAIDA}]{\includegraphics[width=0.45\linewidth]{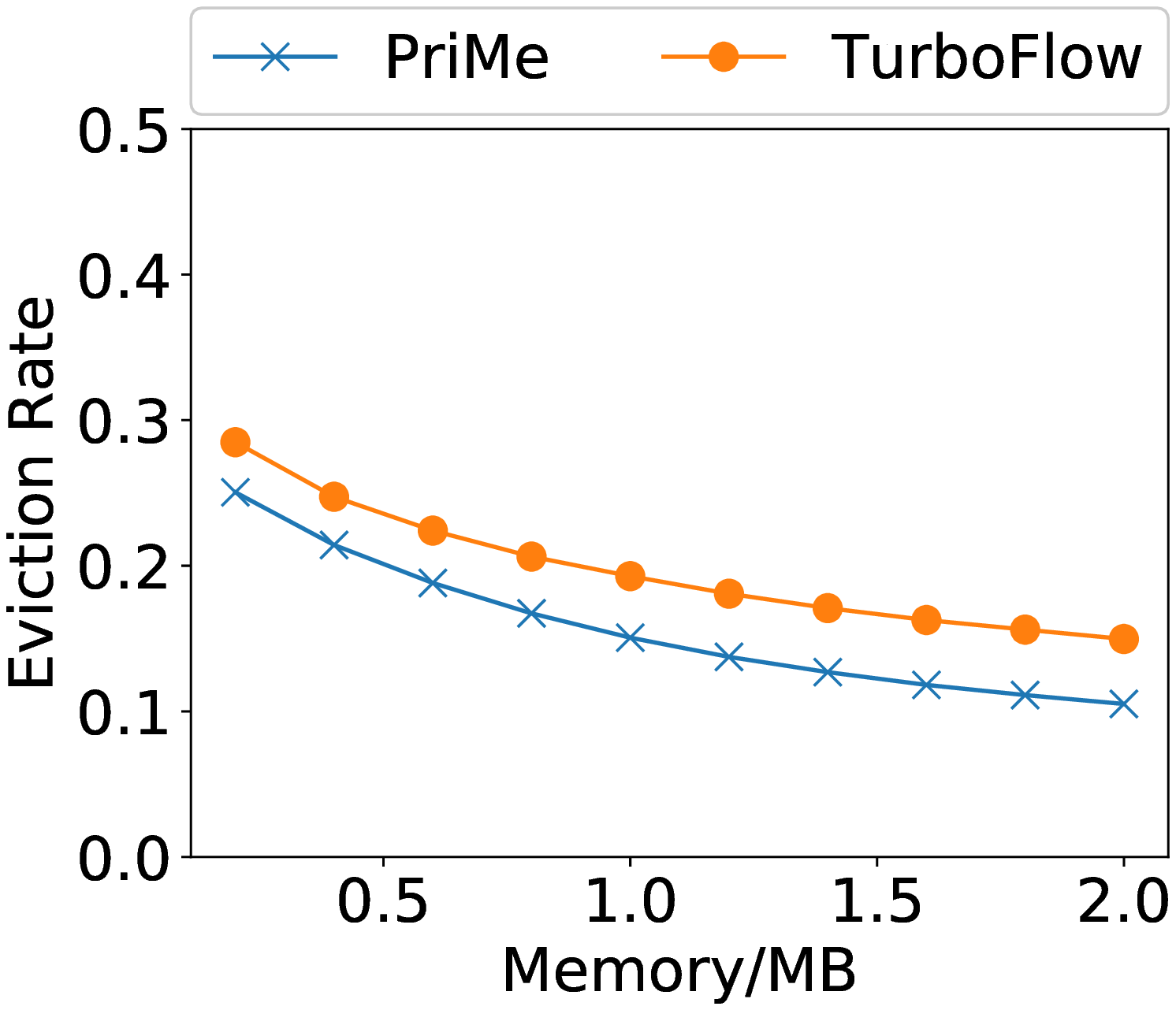}}
		\subfigure[HGC\label{HGC}]{\includegraphics[width=0.45\linewidth]{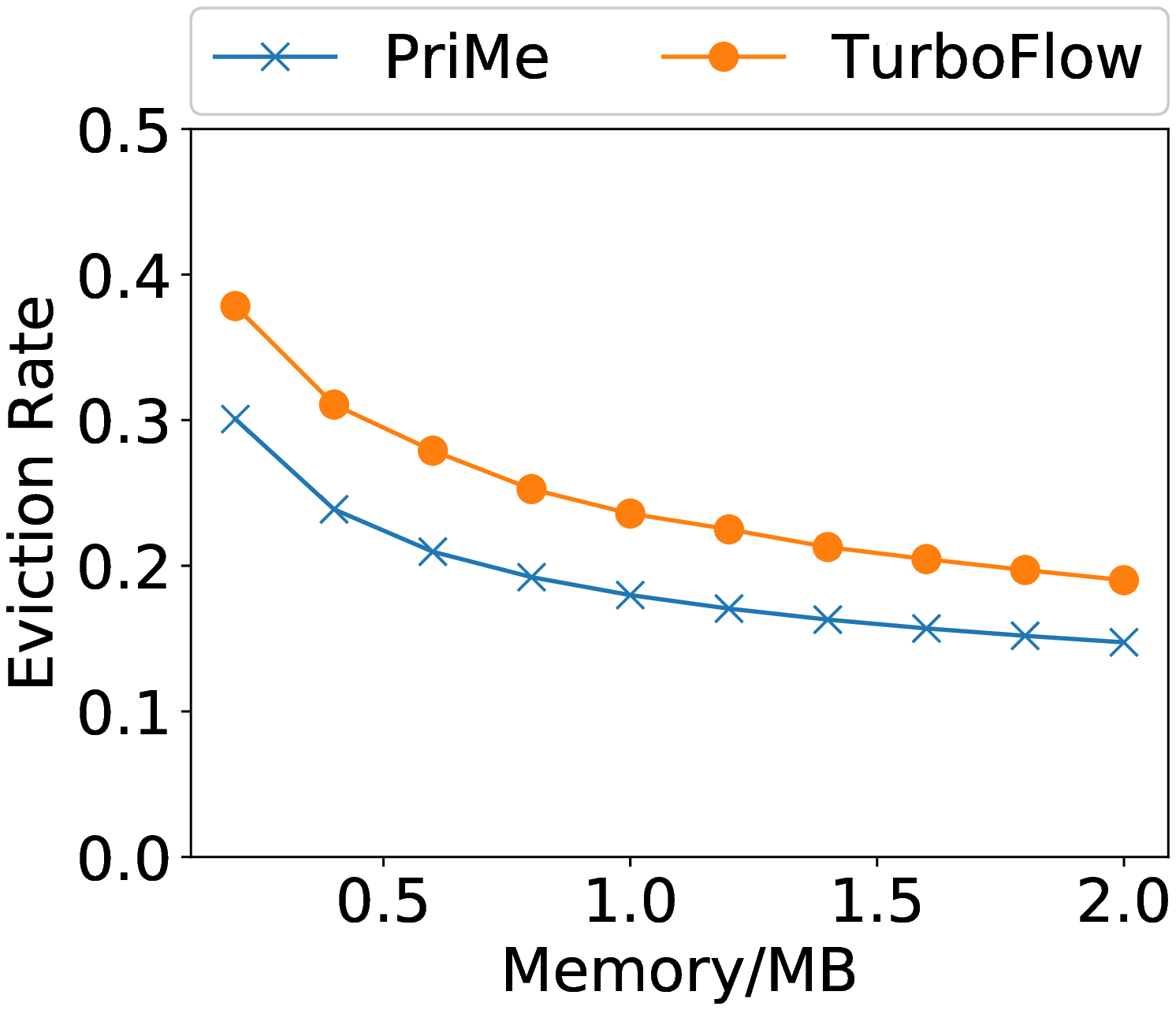}}
	}
	\caption{The collision rates of PriMe and TurboFlow for various memory size.}
	\label{fig:collision_rates_with_various_memory_size}
\end{figure}

Then we choose a trace file from CAIDA and HGC respectively, and increase the memory size from 0.2 MB to 2 MB in the step of 0.2 MB. As shown in Fig.~\ref{fig:collision_rates_with_various_memory_size}, the \emph{eviction rates} of PriMe and TurboFlow decrease constantly as the memory size increases. When the memory size is 2 MB, the \emph{eviction rate} of PriMe is 30\% smaller than that of TurboFlow in CAIDA trace, and it is 23\% smaller than that of TurboFlow in HGC trace.

\subsection{The Performance of DRAM Part}
Since TurboFlow, the only solution available that shares the same target with PriMe, does not optimize the DRAM part, we simply calculate the \emph{aggregation rate} and \emph{redundancy} of PriMe as defined above. We choose a trace file from CAIDA and HGC respectively, and replay 5 million packets from each trace file. 

\begin{figure}[t]
	\centering
	\mbox{
		\subfigure[\label{subfig:aggregation_rate}]{\includegraphics[width=0.45\linewidth]{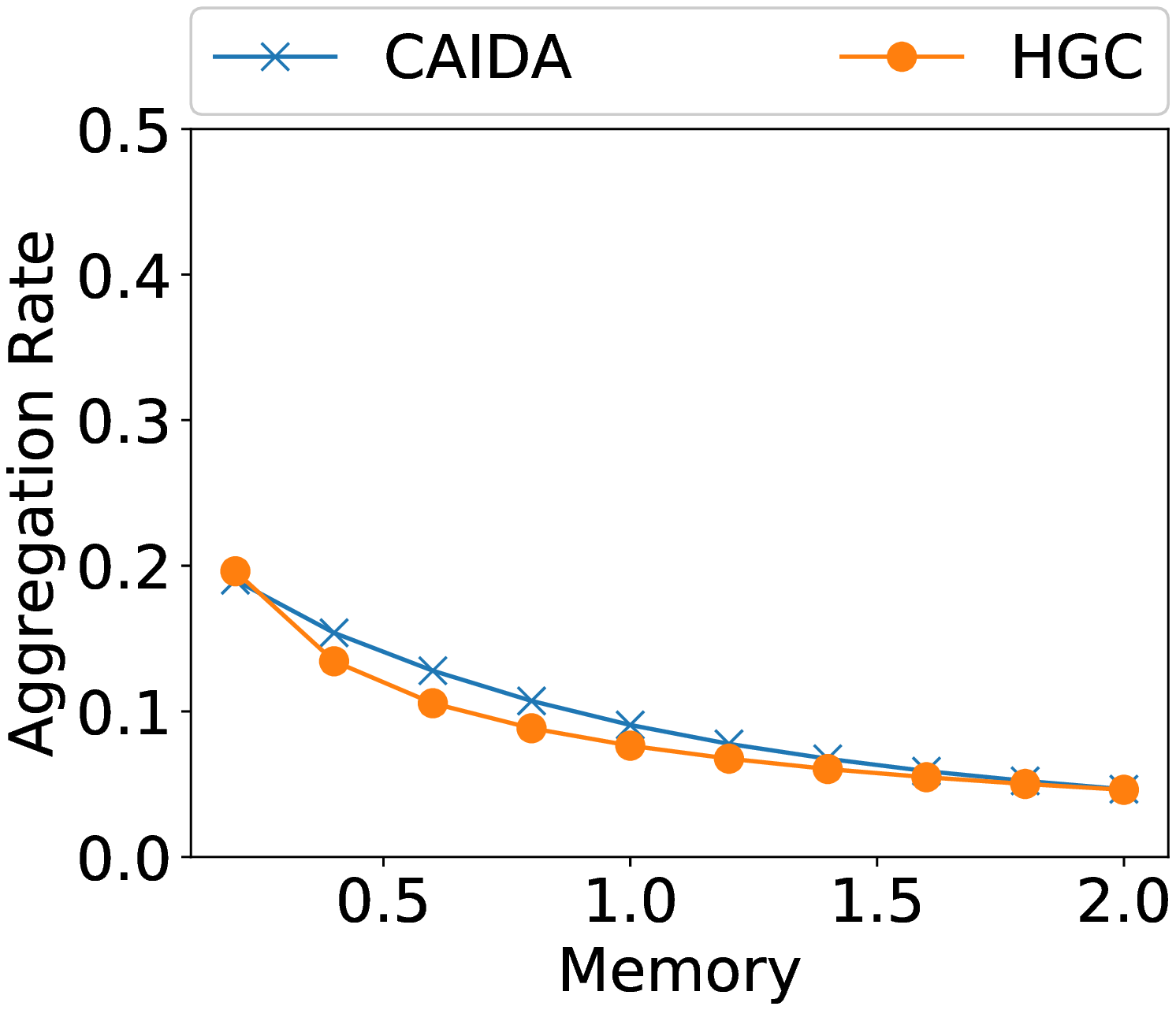}}
		\subfigure[\label{subfig:redundancy}]{\includegraphics[width=0.45\linewidth]{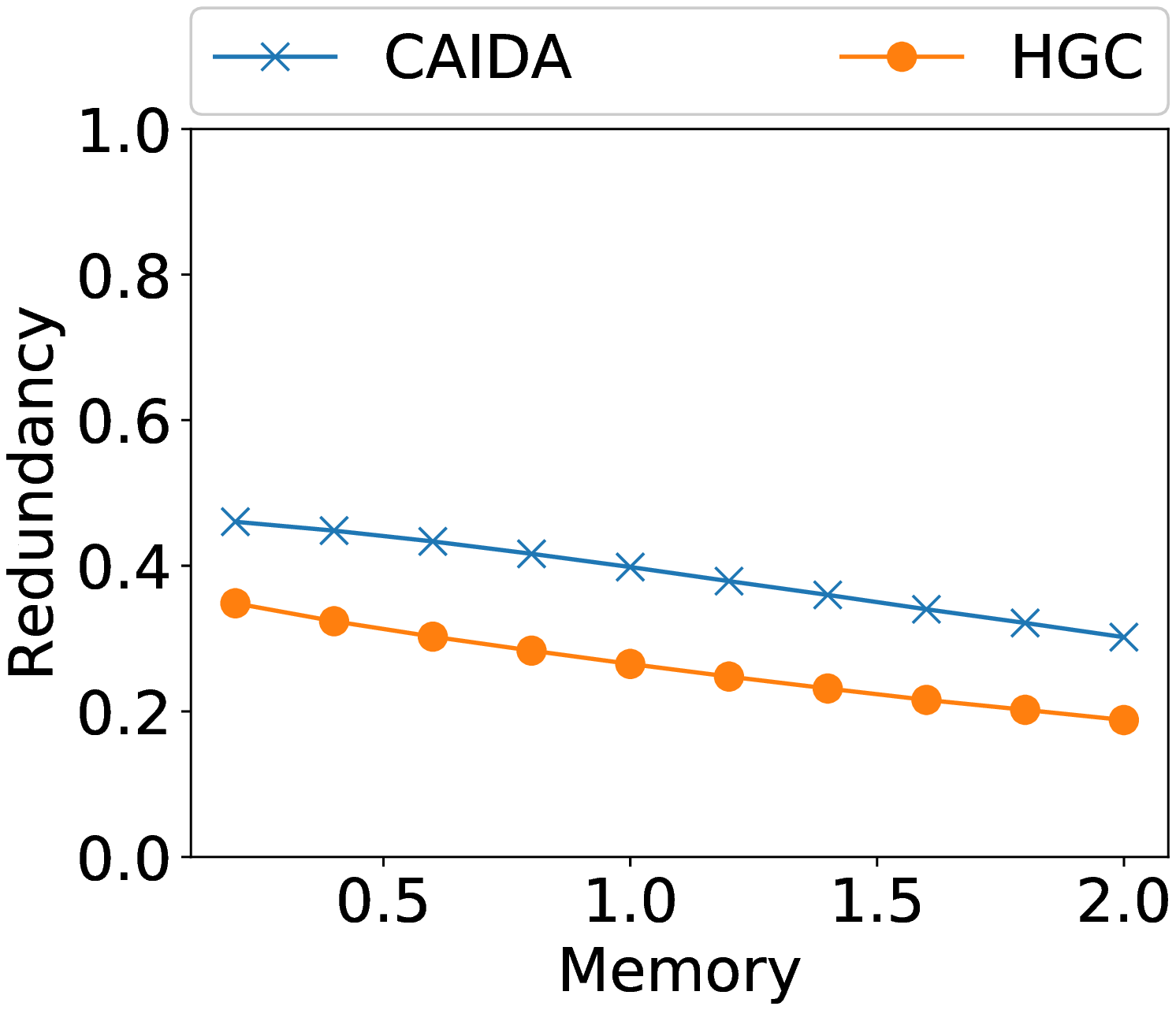}}
	}
	\caption{The \emph{aggregation rate} and \emph{redundancy} of PriMe in CAIDA trace and HGC trace for various memory size.}
	\label{fig:aggre_rate_accuracy_PriMe_various_memory}
\end{figure}

As shown in Fig.~\ref{subfig:aggregation_rate}, when the SRAM size increases from 0.2 MB to 2.0 MB, the \emph{aggregation rate} is decreased from around 19\% to 4.6\% for both CAIDA trace and HGC trace, which means that the \emph{Aggre} need to process 4.6 packets only for every 100 packets processed by ASIC when the size of SRAM allocated is 2.0 MB. As the memory accesses and computations of the DRAM part is mainly from the \emph{Aggre} component, this result implies that the DRAM part can keep up with the SRAM part even if DRAM is 20$\times$ slower than SRAM.

Fig.~\ref{subfig:redundancy} shows that as the SRAM size is increased from 0.2 MB to 2.0 MB, the \emph{redundancy} of PriMe is decreased from 46\% to 30\% in CAIDA trace and from 35\% to 19\% in HGC trace, which means only 19\% of the resulting TFRs are redundant in the best case, and there will be fewer redundant TFRs if more SRAM is allocated.

\section{Conclusion}
Our target in this work is to propose a per-flow network measurement solution that can record each flow accurately, by combining the small-amount but fast SRAM and the large-amount but slow DRAM. Primary simulation shows that our algorithm, PriMe, can do this, as it allows the difference of access time between SRAM and DRAM to be as large as 20 times, or ever more. However, our solution is not perfect, as multiple TFRs may exist for a single flow, and nearly half of the resulting TFRs are redundant in the worst case and 19\% of the resulting TFRs are redundant in the best case.
\bibliographystyle{IEEEtran}
\bibliography{ms}
\end{document}